\documentclass[12pt,preprint]{aastex}

\usepackage{textcomp}
\usepackage{amsmath}
\usepackage{graphicx}
\usepackage{natbib}

\shorttitle{Do IMBHs exist in globular clusters?}
\shortauthors{Sun et al.}

\begin{document}

\title{Do Intermediate-Mass Black Holes Exist in Globular Clusters?}

\author{Mou-Yuan Sun\altaffilmark{1},
Ya-Ling Jin\altaffilmark{1},
Wei-Min Gu\altaffilmark{1},
Tong Liu\altaffilmark{1},
Da-Bin Lin\altaffilmark{1,2},
and Ju-Fu Lu\altaffilmark{1}}

\altaffiltext{1}{Department of Astronomy and Institute of Theoretical Physics
and Astrophysics, Xiamen University, Xiamen, Fujian 361005, China;
guwm@xmu.edu.cn}
\altaffiltext{2}{Department of Physics and GXU-NAOC Center for Astrophysics
and Space Sciences, Guangxi University, Nanning 530004, China}

\begin{abstract}
The existence of intermediate-mass black holes (IMBHs) in globular clusters (GCs)
remains a crucial problem. Searching IMBHs in GCs reveals a discrepancy between
radio observations and dynamical modelings: the upper mass limits constrained by
radio observations are systematically lower than that of dynamical modelings. One
possibility for such a discrepancy is that, as we suggest in this work, there
exist outflows in accretion flows. Our results indicate that, for most
sources, current radio observations cannot rule out the possibility that IMBHs
may exist in GCs. In addition, we adopt an $\dot {M}-L_{\rm R}$ relation
to revisit this issue, which confirms the results obtained by the Fundamental
Plane relation.
\end{abstract}


\section{Introduction}
Over the past decades, astronomers have suspected that there might be a population
of intermediate-mass black holes (IMBHs), which connect the stellar-mass black
holes and the supermassive black holes. On the observational side, accretion of IMBHs
could be the energy source of some hyper-luminous X-ray sources (HLXs), e.g., the
well-known HLX-1 \citep{hlx1, dhlx11}. On the theoretical side, IMBHs are expected to
be formed in centers of globular clusters (GCs) \citep[e.g.][]{mh02}. Some of these
IMBHs can be perfect candidates of first black hole seeds which eventually grow up to
be SMBHs via merge or accretion \citep{vol12}. Others, however, may survive today. It
is meaningful to search IMBHs in GCs.

One way to search IMBHs in GCs is to apply a dynamical modeling: model kinematic 
data of GCs. Many works have been done with this method (for the results of some 
GCs, see Table~1). These works suggested that GCs harbor IMBHs
($10^2 \sim 10^4 M_{\odot}$) in their centers. Another way to test the 
existence of IMBHs in GCs is to detect the unique signatures emitted by surrounding 
accretion flows. Unfortunately, the accretion rates to IMBHs (if IMBHs do exist) are 
expected to be extremely low since GCs have few gases. Therefore, it is quite 
difficult to directly detect X-ray emission. \cite{mradio04} suggested that radio 
observation could be a promising tool to probe IMBHs in GCs. The idea is 
based on the scenario that there exists a universal correlation among the X-ray 
luminosity ($L_{\rm{X}}$), the radio luminosity ($L_{\rm{R}}$), and the mass of 
black hole ($M_{\rm BH}$) \citep[e.g.,][see also in Section~2.1]{m03,fkm04,wwk,plotkin}. 
Basically, this so-called Fundamental Plane relation suggests that the radio/X-ray 
ratio increases with $M_{\rm BH}$. One can constrain the mass of an IMBH by using 
the Fundamental Plane relation, radio observations and the accretion theory (for the 
details, see Section 2). Also, the upper mass limits of IMBHs in GCs can be settled if 
radio observations do not detect any structures within the sensitivity of radio telescopes. 
Recently, \cite{strader12} obtained the $3\sigma$ upper limits of radio luminosity in 
three GCs (M15, M19, M22) with JVLA observations. Based on these data, they found that 
the corresponding upper limits of $M_{\rm BH}$ are too small for these GCs to harbor any 
IMBHs. Moreover, radio observations for other sources (see Table~1) also showed conflicts 
between the resulting upper limits of $M_{\rm BH}$ and the mass constrained via dynamical 
modelings \citep[for details, see discussions in][and our Figure~1]{strader12}. This 
discrepancy indicates that either IMBHs do not exist in GCs or the accretion onto 
IMBHs is significantly weaker than that predicted by the theory.

In this work, we take the role of outflows into consideration. As a consequence, 
the accretion rate onto IMBHs (and $L_{\rm R}$) will be significantly lower than 
what previously predicted. On the other hand, all sources in Table~1 seem to be in the 
quiescent state. The Fundamental Plane relation, as argued by \cite{yc05}, should 
steepen into a new relation which differs from the one suggested by, for example, 
\cite{m03}. Moreover, the Fundamental Plane relation, as mentioned by \cite{plotkin}, 
seems to indicate that the X-ray emission should originate from the jet rather than the 
ADAF. Considering that the X-ray luminosity in this work and some previous works is 
estimated by the ADAF solution, we therefore explore the validity of using the 
Fundamental Plane relation and the robustness of these results. The paper is 
organized as follows. In Section~2, we briefly summarize previous works on this 
issue. In Section~3, we compare our results with radio observations. Summary and 
discussion are given in Section~4.

\section{Methods in previous works}
\label{pw}
In this section, we will discuss physics of constraining mass of IMBHs from radio 
observations in a detailed way \citep[see also,][]{mradio04}.
The first step is on the calculation of $\dot{M}$. Previous works usually assume 
that the accretion rate is a fraction of Bondi accretion rate, i.e., 
$\dot{M}=f\dot{M}_{\rm{B}}$ \citep[e.g., in][$f=0.03$]{strader12}. The Bondi 
accretion is the spherical accretion starting from the Bondi radius 
$R_{\rm B} = 2GM_{\rm BH}/c_{\rm s}^2(\infty)$, where 
$c_{\rm s}=(\gamma k_{\rm B}T/\mu m_{\rm{H}})^{1/2}$
is the speed of sound, and $\gamma$, $k_{\rm B}$, $T$ and $\mu$ are the ratio of 
specific heats, the Boltzmann constant, the temperature of gases and the mass per 
particle, respectively. Since the typical temperature of gases in GCs is around 
$T=10^4 {\rm K}$, we have $R_{\rm B} \approx 10^9 T_4^{-1}R_{\rm s}$, where 
$T_4 = T/(10^4 {\rm K})$, and $R_{\rm s} \equiv 2GM_{\rm BH}/c^2$ is the 
Schwarzschild radius. The corresponding Bondi accretion rate is \citep{book}
\begin{equation}
\label{mb} 
\dot{M}_{\rm{B}}=(\frac{2}{5-3\gamma})^{\frac{5-3\gamma}{2(\gamma - 1)}} 
\pi G^2M_{\rm BH}^2\frac{\rho(\infty)}{c_{\rm{s}}^3(\infty)} \ ,
\end{equation} 
where $\rho$ is the mass density of gases (in this work, we adopt $\rho=0.2
m_{\rm H}$ cm$^{-3}$, $\gamma=1.4$, and $\mu=1.23$).

By knowing the accretion rate, one can now estimate $L_{\rm{X}}$ according to the 
accretion theory. The Eddington ratio of accretion rates is 
$\dot{m} \equiv \dot{M}/ \dot{M}_{\rm Edd} \approx 4.3\times 10^{-5} f m_{2\rm k}$, 
where the Eddington rate is defined as $\dot{M}_{\rm Edd} =10 L_{\rm Edd}/c^2$, 
and $m_{2\rm k} = M_{\rm BH}/(2000 M_{\odot})$. Therefore, the flow should be an 
advection-dominated accretion flow (so-called ``ADAF''). The corresponding 
radiative efficiency $\eta$ usually (although very roughly) scales as 
$\eta=0.1\dot{m}/\dot{m}_{\rm{c}}$, where near the IMBH $\dot{m}_{\rm{c}}\approx 
0.01$ is the critical accretion rate of ADAFs. The 
bolometric luminosity is calculated by 
\begin{equation}
L_{\rm{Bol}}=\eta \dot{M}c^2 \ .
\end{equation}
Previous works usually use $L_{\rm{X}}=L_{\rm{Bol}}$. Clearly, both of them are 
a function of the mass of IMBHs.

Since $L_{\rm{X}}$ scales as a function of $M_{\rm BH}$, one can obtain a
one-to-one relation between $M_{\rm BH}$ and $L_{\rm R}$ by adopting the 
Fundamental Plane relation, 
\begin{equation}
\label{fp}
\log L_{\rm{X}} =A\log L_{\rm{R}} +B\log M_{\rm{BH}}+C \ ,
\end{equation}
where $L_{\rm{X}}$ and $L_{\rm{R}}$ are both in units of $\rm erg~s^{-1}$, and 
$M_{\rm{BH}}$ is in units of the solar mass. The parameters $A$, $B$ and $C$ are 
constrained by observations. Equations~(\ref{mb})-(\ref{fp}) reveal that the 
non-detection of radio signals in GCs provide an upper limit of $M_{\rm{BH}}$.

However, attentions should be paid on this method. The first thing is about 
$\dot{M}$ near IMBHs, where most of energy is released. On one hand, since 
$\dot{m}_{\rm{c}}\sim 10^{-5}$ at $R_{\rm{B}}$ \citep[$\dot{m}_{\rm{c}}$ scales 
as $\sim 0.01 (r/10^3)^{-1/2}$, see e.g.,][where $r=R/R_{\rm{s}}$]{na98} and the 
angular momentum of gases in GCs is low, ADAFs are likely to extend to $R_{\rm{B}}$. 
When the angular momentum of gases in ADAFs is considered, the actual accretion 
rate at $R_{\rm B}$ is only a fraction of $\dot{M}_{\rm{B}}$, i.e., 
$\dot{M}(R_{\rm B})=\alpha \dot{M}_{\rm{B}}$ \citep{na98}, 
where $\alpha$ is the viscosity parameter. On the other hand, ADAFs are likely to 
suffer significant outflows \citep[e.g.,][]{qn99, lc09, cao10, yuan12}. Following 
\cite{yuan12}, outflows can be described by the following assumption: 
\begin{equation}
\label{adaf}
\dot{M}=\alpha \dot{M}_{\rm B}(\frac{R}{R_{\rm out}})^s \ ,
\end{equation}
where $R_{\rm out}$ is the outer radius of the ADAF. In this work we assume 
$R_{\rm{out}}=R_{\rm{B}}$. The parameter $s$, which determines the strength of 
outflows in ADAFs, cannot be self-determined by theoretical considerations. 
However, as pointed out by \cite{yuan12}, some simulations (both hydrodynamical 
and magnetic-hydrodynamical) indicate $s = 0.4\sim 0.5$ \citep[observations of 
NGC 3115 also suggest $s\approx 0.4\sim 0.5$, see][]{wong11}. Therefore, the 
accretion rate near the IMBH is $\sim (10/R_{\rm{B}})^{0.4} \dot{M}_{\rm{out}} 
\sim 6\times10^{-4}\alpha \dot{M}_{\rm{B}}$, which is smaller than 
that of \cite{strader12} (who assumed that $0.03 \dot{M}_{\rm{B}}$ are accreted 
onto IMBHs).

Another thing is on the estimation of $\eta$, especially when ADAFs suffer 
outflows. Recently, \cite{xy12} have done a detailed calculation on $\eta$, which 
included the effects of outflows and found
\begin{equation}
\label{rf}
\eta \equiv L_{\rm{Bol}}/\dot{M}_{\rm net}c^2
= \eta_0 (\frac{\dot{m}_{\rm net}}{\dot{m}_{\rm c}})^{\beta} \ ,
\end{equation}
where $\dot{M}_{\rm net}$ is the accretion rate near the horizon of the black hole. 
The parameters $\eta_0$ and $\beta$ depend on the fraction of energy that directly 
heats electrons (denoted by $\delta$). For $\dot{m}$ less than $\sim 10^{-4}$ and 
$\delta =0.001~(0.1)$, \cite{xy12} shows $\eta_0 = 0.065~(0.12)$ and 
$\beta = 0.71~(0.59)$. 

As seen from Equation~(\ref{fp}), $L_{\rm{R}}$ is sensitive to $L_{\rm{X}}$. 
The usually adopted $L_{\rm{X}}$ in fitting the Fundamental Plane 
relation is in the range of $1\sim 10$~keV \citep[e.g.][]{m03,plotkin}, which is 
only a fraction of the bolometric luminosity. This is because ADAF is optically 
thin and the spectrum is not a blackbody type but roughly a flat spectrum over 
$\sim 10$ orders of magnitude (from radio to hard X-ray) in the $\nu-\nu L_{\nu}$ 
diagram \citep[e.g.,][]{qn99}. It is therefore reasonable to assume
\begin{equation}
\label{lx}
L_{\rm{X}} (1-10\ {\rm keV}) = \zeta L_{\rm bol} \ ,
\end{equation}
with $\zeta \sim 0.1$. In our opinion, $L_{\rm{X}}$ in some previous works is 
overestimated.

So far, we have argued that some modifications are required in determining mass 
of IMBHs by radio observations. In the following section, we will apply these 
modifications and present our results.

\section{The existence of IMBHs in GCs}
\subsection{The predicted radio luminosity}
\label{lrsb}
Applying all the modifications discussed in Section~2, we can derive a
new relation between $L_{\rm{R}}$ and $M_{\rm{BH}}$ with 
Equations~(\ref{fp})-(\ref{lx}) and the definition of $\dot{M}_{\rm{B}}$
(Equation~\ref{mb}),
\begin{equation}
\label{re}
\begin{split}
\log L_{\rm{R}} & = \frac{38.05-3.30B-C-2.37\beta+\log (\zeta \eta_0)}{A}\\& 
+\frac{(\beta+1)\log(\lambda)}{A}+\frac{\beta+2-B}{A}\log(m_{2\rm{k}}) \ ,
\end{split}
\end{equation}
where $\lambda=\alpha(R_{\rm{in}}/R_{\rm{B}})^s$ is the ratio between 
$\dot{M}_{\rm net}$ and $\dot{M}_{\rm{B}}$, and $R_{\rm{in}}$ is the inner radius 
of outflows. Following \cite{yuan12}, we adopt $s=0.4$ and $R_{\rm{in}}=10 R_{\rm{s}}$. 
We assume $\alpha=0.1$ and $\zeta=0.1$. For the Fundamental Plane relation, we 
adopt the one fitted by \cite{plotkin}, i.e., $A=1.45$, $B=-0.88$, and $C=-6.07$ 
\citep[We choose this version of Fundamental Plane relation due to
the following two reasons. 
First, the sample corresponding to this relation consists of sources with flat/inverted
radio spectrum. When considering the radio observations of globular clusters, we usually 
assume a flat radio spectrum. Second, this sample, as pointed out by][
minimizes the systematical bias of synchrotron cooling and thus can be considered as 
the most robust relation.]{plotkin}. As for $\delta$, we choose two typical values 
$\delta=0.1$ and $\delta=0.001$. Note that with $\lambda=0.03$, $\beta=1$, $\zeta=1$, 
and $\eta_0=0.1$, Equation~(\ref{re}) can recover the results of \cite{strader12}.

Now, we can use Equation~(\ref{re}) to testify whether the recent radio non-detection 
results can rule out the existence of IMBHs in GCs or not. Table~1 lists the 
information of some GCs which may harbor IMBHs. $L_{\rm{R}}$ listed in Table~1 is 
the $3\sigma$ upper limit of radio luminosity constrained by radio observations. We 
calculate the predicted $L_{\rm{R}}$ for each source by using Equation~(\ref{re}) 
and mass listed in Table~1, and compare them with radio observations. The mass of 
IMBHs in Table~1 was obtained by dynamical modelings.

Figure~1 plots $L_{\rm R}$ as a function of $M_{\rm BH}$. As shown by Figure~1, 
$L_{\rm{R}}$ in our results are significantly lower than that of \cite{strader12}. 
The major reason is relevant to outflows. The actual accretion rates near IMBHs are 
lower than that of \cite{strader12}. More importantly, for most sources $L_{\rm{R}}$ 
in our results are obviously smaller than the $3\sigma$ upper limit of $L_{\rm{R}}$ 
given by radio observations. Thus, our results indicate that the current radio 
observations seem not to conflict with the dynamical modelings. In other words, 
IMBHs may still exist in GCs. An interesting exception is $\omega$ Cen. As seen 
from Figure~1, even in the presence of outflows, the predicted $L_{\rm{R}}$ is 
close to the $3\sigma$ upper limit of $L_{\rm{R}}$. Therefore, $\omega$ Cen is 
unlikely to harbor a $\sim 10^4 M_{\odot}$ IMBH (see the last section for 
more details). In addition, the thin solid line and the dotted line in 
Figure~1 represents $L_{\rm R}$ estimated by the $\dot{M}-L_{\rm R}$ relation 
instead of the Fundamental Plane relation. We will interpret these results in the 
following subsection.

There are two issues we have to mention. The first issue is about the relatively 
large scatter in the Fundamental Plane relation. We will discuss this problem in the 
last section. The second one is the consistency of the assumptions that adopted 
in such an issue and previous works. The Fundamental Plane relation have been 
explored by many works. For instance, \cite{yc05} suggested that this relation can be 
explained in the framework that X-ray emission is predominantly from ADAFs whereas 
the radio emission is from jets. Moreover, they argued that in the extremely low 
luminosity region this relation should break into a new one. The argument is that, 
below a critical X-ray luminosity, $L_{\rm X,c}\approx 10^{-5}-10^{-6}L_{\rm Edd}$, 
the X-ray emission from the jet (radiatively cooled) should dominate over that from 
ADAFs (so-called ``quiescent state"). In this spirit, all the sources in Table~1 should 
locate in this quiescent state. It seems more appropriate to use the Fundamental Plane 
relation obtained by \cite{yc05}. Note that there are two Galactic black hole 
X-ray binaries (BHXRBs), A0620-00 \citep{G06} and V404 Cyg \citep{C08}, whose X-ray 
luminosity is well below the critical X-ray luminosity proposed by \cite{yc05}. The 
corresponding radio/X-ray correlation analyses are, however, inconsistent with 
the Fundamental Plane relation of \cite{yc05}. In our opinion, there are
two possibilities for these inconsistent results. The first one may be related to a large scatter 
in determining $L_{\rm X,c}$ (at least for V404), since $L_{\rm X,c}$ was 
constrained by assuming that there is a smooth transition of the Fundamental Plane relation 
from the type of \cite{m03} to that of \cite{yc05} at $L_{\rm X,c}$. Actually, if
the Fundamental Plane relation of \cite{plotkin} and that of \cite{yuan09} are used
to constrain $L_{\rm X,c}$, one will find $L_{\rm X,c}\sim 10^{-7}L_{\rm Edd}$. The second 
possibility is related to the cooling of the jet. As mentioned by \cite{yc05}, their 
Fundamental Plane relation is based on the assumption that
the jet should be radiatively cooled in the X-ray bands. However, for the above
two sources, such an assumption may be invalid due to the following reason.
According to the synchrotron cooling frequency (under such circumstance,
the Compton scattering can be neglected)
$\nu_{\rm break}\propto m^{-1/2}\dot{m}^{-3/2}$ \citep[e.g.,][]{h04}, 
for low mass black holes and low accretion rates, the jet of the BHXRB may be uncooled.

On the other hand, at higher X-ray luminosity, the Fundamental 
Plane relation of \cite{plotkin} indicates that the X-ray emission should be dominated
by the synchrotron emission of the uncooled jet rather than the ADAF \citep[see Figure~5 of ][]{plotkin}. 
Thus, both the arguments of \cite{yc05} and \cite{plotkin} imply that the results 
of Section~\ref{lrsb} \citep[and the results of some previous works, e.g., ][]{strader12} would 
be suspicious according to the fact that the X-ray emission we considered in 
Section~\ref{lrsb} is from ADAFs rather than jets.
We will address this issue in the following subsection.

\subsection{The influences of the Fundamental Plane relation and X-ray processes on 
the estimation of $M_{\rm{BH}}$ from $L_{\rm{R}}$}

As stated in Section 3.1, there exists inconsistency between the Fundamental Plane relation 
we adopted and the radiative processes of X-ray emission we assumed. Then, it is 
natural to ask whether the results obtained in Section~\ref{lrsb} will change significantly 
if we instead calculate the X-ray emission of the radiatively cooled jet and use the 
Fundamental Plane relation of \cite{yc05}, or obtain the X-ray emission of the uncooled 
jet and use the Fundamental Plane relation of \cite{plotkin}. However, it is not easy to 
constrain the X-ray emission of the jet (whether cooled or uncooled) due to our poor 
knowledge of the jet physics. Therefore, we will investigate the issue in a different way, which 
is based on the idea that there may exist a correlation between $\dot{M}$ and the radio 
luminosity for flat spectrum radio cores \citep[e.g.,][]{BK79, hs03, k06}. The key point is 
whether this $\dot{M}-L_{\rm{R}}$ relation, as first quantitatively obtained by \cite{k06} 
for low luminosity BHXRBs, depend on the origin of X-ray emission. If the sources
whose X-ray emission is dominated by ADAFs share the similar relation with the sources whose 
X-ray emission mainly comes from jets, then one can expect that the results of Section~\ref{lrsb} 
is robust. The physical reason is as follows. In Section~\ref{pw}, $L_{\rm{X}}$ is obtained 
by $L_{\rm{X}}=\zeta \eta \dot{M}_{\rm net}c^2$ (see Equation~\ref{lx}) rather than via 
X-ray observations. This equation and the Fundamental Plane relation actually reveal a 
direct relation between $\dot{M}$ and $L_{\rm{R}}$. In other words, Equation~(\ref{re}) 
is identical to an $\dot{M}-L_{\rm{R}}$ relation. Below, we will explore the 
$\dot{M}-L_{\rm{R}}$ relation of both ADAF-dominated and jet-dominated sources.

To answer this question, we search the literature for published information on $\dot{M}$ 
and $L_{\rm{R}}$. Our sample is collected from \cite{wu07}(seven FR I galaxies), 
\cite{yuan05}(XTE J1118+480), \cite{yuan09}(14 LLAGN\footnote{Note that M32 is excluded 
because only an upper limit of $L_{\rm R}$ is obtained; M87 \citep[jet-dominated 
according to][]{yuan09} is also excluded because currently the origin of the X-ray emission
is still under debate, especially the numerical simulation of \cite{HL12} suggests that the 
ADAF can account for the X-ray emission.}) and \cite{zhangh10}(three BHXBs) (see Table~2). 
Accretion rates of these sources are obtained by fitting the coupled ADAF-Jet model 
\citep[for readers who are interested in details of this model, we recommend][]{yc05} to 
the overall spectral energy distribution (SED) of each source. Below we will try 
to summarize the main assumption of the coupled ADAF-Jet model.

The accretion flow is described by an ADAF with outflows (i.e., Equation~\ref{adaf}). 
To fully account for the global solution of the ADAF, one should specify $\dot{M}_{\rm out}$, 
$R_{\rm out}$, $s$, the viscosity parameter $\alpha$, and the magnetic 
parameter $\beta$. The SED of the flow can be obtained
after the global solution is solved \citep[see e.g.,][]{qn99}. 
On the other hand, the jet model is quantified based on the internal shock scenario used 
in gamma-ray burst models. In this model, a fixed fraction of material of the flow is lost 
into form a jet ($\dot{M}_{\rm jet}$). Other two parameters that quantify the geometry and 
motion of the jet are the half-opening angle $\phi$ and bulk Lorentz factor $\Gamma_{\rm jet}$. 
According to the internal shock scenario, shocks occur as shells with different velocity 
colliding with each other. As a consequence, a few fraction of electrons in the jet is 
accelerated into a power-law distribution (i.e., $N(E)\propto E^{-p}$, typically, $p\sim 2$). 
The remaining parameters are the fraction of accelerated electrons $\xi_{\rm e}$ and magnetic 
field in the shock front $\xi_{\rm B}$. With these parameters, one can calculate the SED of 
the jet by considering the synchrotron emission and/or Compton scattering. Note that
the radiative cooling of the non-thermal electrons
is also taken into consideration for the X-ray emission of the jet.
Comparing the SED of coupled ADAF-Jet model with
observations, one can, in principle, constrain $\dot{M}_{\rm out}$ (or $\dot{M}_{\rm jet}$ 
if X-ray emission is dominated by the jet), $R_{\rm out}$ \citep[for an interesting discussion 
of the robustness of the obtained parameters, see Figures~4-6 of][which indicate that the 
calculated SED is inconsistent with data if $\dot{M}$ varies a little 
from the best fitting value]{zhangh10}. Besides, the origin of the X-ray emission can be 
found (indicated in Table~2).

We also include the data of BHXRBs used in \cite{k06}. These BHXRBs are under the 
state transition from the soft state to the hard state (except for GRS~1915+105, which is in 
a so-called ``plateau state"). $\dot{M}$ can be calculated as $\dot{M}=L_{\rm X}/(0.1c^2)$, 
where $c$ is the speed of light. Therefore, as shown by Table~2, we have 22 sources with 
X-ray emission dominated by the ADAF and nine sources with X-ray emission dominated by the 
jet. For the former 22 sources, we perform the 
OLS regression between $\dot{M}$ and $L_{\rm R}$. For the latter nine 
sources, however, both the X-ray emission and the radio emission are dominated by the 
jet, so the obtained $\dot{M}_{\rm out}$ is not reliable. We instead performed the OLS 
regression between $\dot{M}_{\rm jet}$ and $L_{\rm R}$. Our OLS regression results are 
presented in Table~3.

As shown in Table~3, the slope of jet-dominated sources and ADAF-dominated sources 
are consistent with each other under $1\sigma$ uncertainties. However, it is inappropriate to directly 
compare the normalization of jet-dominated sources with that of ADAF-dominated ones 
because $\dot{M}_{\rm jet}$ is only a small fraction of $\dot{M}$. 
We assume $\dot{M}_{\rm jet}=f_{\rm jet}\dot{M}$ and choose a typical value
$f_{\rm jet}=0.05$ to address this issue. By doing this, one can obtain 
an $\dot{M}-L_{\rm R}$ relation for the nine jet-dominated sources. We compare it with the one 
obtained from the ADAF-dominated ones. Figure~2 plots the results. As seen from this figure, 
the $\dot{M}-L_{\rm R}$ relation of jet-dominated sources is consistent with that of 
ADAF-dominated sources under $2\sigma$ uncertainties (filled regions in the plot).
In order to further confirm our results,
we have also tried to take the black hole mass into consideration, 
that is, we perform the OLS multivariate regression among $L_{\rm R}$, $\dot{M}$ 
($\dot{M}_{\rm jet}$), and $M_{\rm BH}$. We find that the requirement of the parameter 
$M_{\rm BH}$ is statistically rejected under the $p$ value is 0.01. Thus, we can
conclude that black holes in low activity state share a similar $\dot{M}-L_{\rm{R}}$ 
relation, regardless of $L_{\rm X}$ or the origin of the X-ray emission.

Since the $\dot{M}-L_{\rm{R}}$ relation does not depend on the origin of the X-ray 
emission, the results obtained in Section~\ref{lrsb} are robust even if the X-ray 
emission may come from the synchrotron emission of the cooled \cite[][]{yc05} or 
uncooled \citep[][]{plotkin} jet. Other evidence that may confirm our conclusion is
the thin solid line and the dotted line in Figure~1, which show $L_{\rm{R}}$ estimated by the 
$\dot{M}-L_{\rm{R}}$ relation (that is, the $\dot{M}-L_{\rm{R}}$ relation of the jet-dominated 
sources with, roughly, $L_{\rm{R}}\sim 10^{20}(\dot{M}_{\rm jet}/10^9\rm{g\ s^{-1}})^{1.56}\ \rm{erg\ s^{-1}}$ 
and $\dot{M}_{\rm jet}=0.05\dot{M}$, since X-ray emission of the sources in Table~1
may mainly come from the jet rather than the ADAF).
For the case of \cite{strader12} ($\dot{M}=0.03\dot{M}_{\rm B}$,
dotted line), the radio luminosity estimated from the $\dot{M}-L_{\rm R}$ is close to (within one order of magnitude)  
the radio luminosity obtained by assuming that the X-ray emission is dominated by the ADAF
and using the Fundamental 
Plane relation (i.e., Section~\ref{lrsb}). The same conclusion holds (again within one
order of magnitude) for our case (thin solid line).

\section{Summary and Discussion}

The radio observation is, as first demonstrated by \cite{mradio04}, useful in probing 
IMBHs in GCs. However, the upper mass limits of IMBHs constrained by radio observations 
are significantly smaller than that of dynamical modelings. In this work, we showed that 
this inconsistency can be solved if ADAFs suffer outflows and IMBHs may exist in GCs. We 
also concluded that the results of Section~\ref{lrsb} and previous works do not 
strongly depend on the physical processes of the X-ray emission and the type of Fundamental Plane 
relation, if $L_{\rm X}$ is obtained via the accretion theory.

The remaining question is about the scatter in the estimation. The Fundamental Plane relation 
of \cite{plotkin} with $A=1.45$, $B=-0.88$ and $C=-6.07$ has a scatter of 
$\sigma_{\rm{int}} = 0.07$. In this work, we have adopted this version of Fundamental 
Plane relation (for reasons, see Section~\ref{lrsb}). The uncertainty in estimating of 
$L_{\rm{X}}$ from $\dot{M}$ also contribute to the scatter. However, for almost all sources 
(except for $\omega$ Cen), the $3\sigma$ upper limits of radio luminosity are at least two 
orders of magnitude higher than that of our estimations. Therefore, the robustness of our 
conclusion that current radio observations cannot rule out the existence of IMBHs in GCs 
will not be affected by the above scatter. 

$\omega$ Cen has long been considered as a promising source that may harbor an IMBH in 
its center. For example, the early work of \cite{Noyola08} analyzed the HST ACS image and 
the Gemini GMOS-IFU kinematic data and concluded that $\omega$ Cen hosts an IMBH with 
$M_{\rm BH}=4.0^{0.75}_{-1.0}\times 10^4\ M_{\odot}$. However, \cite{maad10} explored a 
new dataset of HST proper motion and star count and given an upper limit of the mass of 
the IMBH in $\omega$ Cen: $M_{\rm BH}<1.2\times 10^4\ M_{\odot}$ at $1\sigma$ confidence 
(which is the value we adopted in this work). They also found that data can be well fitted
with a $M_{\rm BH}=0.0$ model and the $M_{\rm BH}$ proposed by \cite{Noyola08} is firmly 
ruled out, although the cluster center of \cite{maad10} is $\sim 12^{''}$ away from that of 
\cite{Noyola08}. Interestingly, \cite{Noyola10} used the VLT-FLAMES to obtain the new 
data and found again that there should be an IMBH with $M_{\rm BH}\sim 5\times 10^4\ M_{\odot}$ 
in the center of $\omega$ Cen \citep[note that even for the center of ][they also concluded 
an IMBH with $M_{\rm BH}\sim 3\times 10^4\ M_{\odot}$]{maad10}. A recent work of \cite{Jalali12} 
confirmed the conclusion of \cite{Noyola10}. In this work, $\omega$ Cen is the only one whose 
$3\sigma$ upper limit of $L_{\rm R}$ is close to the predicted radio luminosity even the
outflows have been taken into consideration. To make the $\sim 4\times 10^4\ M_{\odot}$ IMBH compatible 
with the radio observations of $\omega$ Cen, one has to assume that the mass loss effect of 
outflows is larger than the one with $s=0.4$. Such an assumption challenges recent numerical 
simulations \citep[e.g.,][]{yuan12} and observations \citep[e.g.,][]{wong11}. This leads us to 
conclude that the IMBH in $\omega$ Cen, if exists, is likely to be much lighter than 
$1.2\times 10^4\ M_{\odot}$, which is in agreement with \cite{maad10}. More deep radio 
observations are required to draw robust conclusions.

\acknowledgments

We thank Junfeng Wang, Renyi Ma, and Feng Yuan for beneficial discussions,
and the referee for constructive suggestions to improve the manuscript.
This work was supported by the National Basic Research Program (973 Program)
of China under grant 2014CB845800, and the National Natural Science Foundation
of China under grants 11073015, 11103015, 11222328, and 11233006.

\clearpage

\begin{table*}
\center
\caption{Recent Radio Observations of Globular Clusters}
\label{tab 1}
\begin{tabular}{lcccccccccccc}
\tableline
\hline
Sources  & D (kpc) & log $L_{\rm R}$ & Ref. & $M_{\rm dyn} (M_{\odot})$ & Ref.   \\
(1) & (2) & (3) &(4) & (5) & (6)  \\
\hline
NGC~6388   & 13.2 & 28.684 & 1 & $17000\pm9000$ & 2&\\
$\omega$ Cen   & 5.3 & 27.525 & 3 & $<12000$ & 4&\\
NGC~2808   & 9.5 & 28.940 & 5 & $<10000$ & 6&\\
M54   & 26.3 & 29.6 & 7 & 9400 & 8&\\
M62   & 6.8 & 27.997 & 9 & 3000 & 9&\\
M80   & 10 & 28.332 & 9 & 1600 & 9&\\
47 Tur   & 4.5 & 27.684 & 3 & $<1500$ & 10&\\
M15(1)   & 10.3 & 27.672 & 11 & 1000 & 9&\\
M15(2)   & 10.3 & 28.176 & 9 & 1000 & 9&\\
M19      & 8.2  & 27.322 & 11 & 410 & 12&\\
M22      & 3.2  & 26.505 & 11 & 240 & 12&\\
NGC~6397   & 2.7 & 27.972 & 13 & 50 & 14&\\
\tableline
\end{tabular}
\par
\medskip
\begin{minipage}{0.95\linewidth}
Col.~(1): Name of Globular Cluster. Col.~(2): Distance to
the Sun (kpc). Col.~(3) Logarithm of the center radio luminosity
at 5 GHz ($3\sigma$ upper limit one, in units of $\rm{erg\ s^{-1}}$).
Col.~(5): Mass of Black Holes constrained by dynamical modelings.

REFERENCES:
(1) Cseh et al.~2010; (2) L{\"u}tzgendorf et al.~2011;
(3) Lu \& Kong~2011; (4) van der Marel \& Anderson~2010;
(5) Maccarone \& Servillat~2008; (6) L{\"u}tzgendorf et al.~2012;
(7) Wrobel et al.~2011; (8) Ibata et al.~2009;
(9) Bash et al.~2008; (10) McLaughlin et al.~2006;
(11) Strader et al.~2012; (12) Maccarone~2004;
(13) De Rijcke et al.~2006; (14) $M-\sigma$ relation of Tremaine et al.~(2002).

NOTES: 
1. There are two radio observations for M15. Here, M15(1) corresponds to
Strader et al.~(2012), and M15(2) corresponds to Bash et al.~(2008).
2. For sources without mass uncertainties, there are absent of mass uncertainties 
in the referred paper.
\end{minipage}
\end{table*}

\clearpage

\begin{table*}
\center
\caption{Radio Luminosity and Accretion Rate}
\label{tab 2}
\begin{tabular}{lcccccccccccc}
\tableline
\hline
Sources  & log $\dot{M}$ & log $L_{\rm R}$ & 
$M_{\rm BH} (M_{\odot})$ & $L_{2-10\rm keV}/L_{\rm Edd}$& Ref. & X-ray dominated by   
\\  (1) & (2) & (3) &(4) & (5) & (6) & (7)  \\
\hline
IC~4296   & 23.197 & 39.329 & $1.0\times 10^9$ & $1.3\times 10^{-6}$ & 1 & ADAF &\\
NGC~315   & 23.249 & 40.181 & $3.1\times 10^8$ & $1.5\times 10^{-5}$ & 1 & ADAF &\\
NGC~1052   & 22.523 & 39.669 & $1.26\time 10^8$ & $5.9\times 10^{-6}$ & 1 & ADAF &\\
NGC~4203   & 22.324 & 38.314 & $1.0\times 10^7$ & $1.85\times 10^{-5}$ & 1 & ADAF &\\
NGC~4261   & 24.076 & 39.334 & $4.9\times 10^9$ & $4\times 10^{-6}$ & 1 & ADAF &\\
NGC~6251   & 24.189 & 40.258 & $6\times 10^8$ & $5\times 10^{-5}$ & 1 & ADAF &\\
NGC~4579   & 22.980 & 38.453 & $4\times 10^6$ & $3\times 10^{-4}$ & 1 & ADAF &\\
3C~346   & 24.680 & 41.900 & $7.762\times 10^8$ & $1.8\times 10^{-4}$ & 2 & ADAF &\\
3C~31   & 22.513 & 39.460 & $7.762\times 10^7$ & $4.4\times 10^{-6}$ & 2 & ADAF &\\
3C~317   & 23.225  & 40.730 & $6.31\times 10^8$ & $3.4\times 10^{-6}$ & 2 & ADAF &\\
B2~0055~+30   & 23.627  & 40.260 & $1.514\times 10^9$ & $2.4\times 10^{-6}$ & 2 & ADAF &\\
3C~449   & 22.712 & 39.080 & $2.63\times 10^8$ & $8.0\times 10^{-7}$ & 2 & ADAF &\\
XTE~J1118~+480   & 17.509 & 28.977 & 8 & $\sim 10^{-3}$ & 3 & ADAF &\\
Sw~J1753.05~-0127   & 17.681 & 29.136 & 9 & $\sim 5\times 10^{-3}$ & 4 & ADAF &\\
GRO~J1655~-40   & 17.333 & 28.598 & 6.3 & $\sim 10^{-3}$ & 4 & ADAF &\\
XTE~J1720~-318   & 17.160 & 29.205 & 5 & $\sim 2\times 10^{-3}$ & 4 & ADAF &\\
Cyg~X-1   & 17.586 & 29.823 & $14.8^a$ & $\rm state\ transition^f$ & 5 & ADAF &\\
V404   & 17.624 & 30.408 & $9.0^b$ & $\rm state\ transition^f$ & 5 & ADAF &\\
GX~339-4(1)   & 17.846 & 30.776 & $7.5^c$ & $\rm state\ transition^f$ & 5 & ADAF &\\
1859+226   & 18.327 & 30.612 & $>5.42^d$ & $\rm state\ transition^f$ & 5 & ADAF &\\
GX~339-4(2)   & 18.506 & 30.939 & $7.5^c$ & $\rm state\ transition^f$ & 5 & ADAF &\\
GRS~1915+105   & 19.000 & 30.014 & $10.1^e$ & $\rm plateau\ state^f$ & 5 & ADAF &\\
IC~1459   & 21.286 & 39.319 & $2.0\times 10^9$ & $1.7\times 10^{-7}$ & 1 & Jet &\\
M81   & 19.365 & 36.446 & $7.0\times 10^7$ & $2.3\times 10^{-6}$ & 1 & Jet &\\
M84   & 21.247 & 38.483 & $1.6\times 10^9$ & $1.1\times 10^{-8}$ & 1 & Jet &\\
NGC~3998   & 19.976 & 37.930 & $7.0\times 10^8$ & $3\times 10^{-6}$ & 1 & Jet &\\
NGC~4594   & 20.742 & 37.668 & $1.0\times 10^9$ & $1.2\times 10^{-7}$ & 1 & Jet &\\
NGC~4621   & 19.173 & 35.288 & $2.7\times 10^8$ & $1.9\times 10^{-9}$ & 1 & Jet &\\
NGC~4697   & 18.914 & 34.877 & $1.7\times 10^8$ & $1\times 10^{-9}$ & 1 & Jet &\\
B2~0755~+3   & 22.313 & 40.720 & $8.5\times 10^8$ & $5.2\times 10^{-6}$ & 2 & Jet &\\
3C~66B   & 21.980 & 39.970 & $6.9\times 10^8$ & $1\times 10^{-6}$ & 2 & Jet &\\
\tableline
\end{tabular}
\par
\medskip
\begin{minipage}{0.95\linewidth}
REFERENCES:
(1) Yuan et al.~2009;  (2) Wu et al.~2007;
(3) Yuan et al.~2005;  (4) Zhang et al.~2010;  
(5) K{\"o}rding et al.~2006.
\\
NOTES: 
a. Mass adopted from Orosz et al.~(2011); 
b. Mass adopted from Khargharia et al.~(2010); 
c. Mass adopted from Chen~(2011); 
d. Mass adopted from Corral-Santana et al.~(2011); 
e. Mass adopted from Steeghs et al.(2013); 
f. Sources in the hard state but close to the state transition (GRS~1915 is in 
the radio plateau state); 
g. For the nine jet-dominated sources, $\dot{M}$ represents jet mass loss rate
rather than accretion rate.

\end{minipage}
\end{table*}

\clearpage

\begin{table*}
\center
\caption{The Fitting Results}
\label{tab 3}
\begin{tabular}{lcccccccccccc}
\tableline
\hline
X-ray emission dominated by & a & b & $p$ value  \\
(1) & (2) & (3) &(4)  \\
\hline
ADAF   & $-15.42\pm 0.72$ & $1.76\pm 0.06$ & $<2.2\times 10^{-16}$ &\\
Jet   & $-10.22\pm 1.78$ & $1.56\pm 0.15$ & $1.89\times 10^{-5}$ &\\
\tableline
\end{tabular}
\par
\medskip
\begin{minipage}{0.95\linewidth}
NOTES: 
1. For sources whose X-ray emission is dominated by the ADAF, the fitting relation is 
$\log(L_{\rm R}/10^{30}\ \rm{erg\ s^{-1}})=a+b*\log(\dot{m}/10^9\ \rm{g\ s^{-1}})$; Otherwise, 
the fitting relation is $\log(L_{\rm R}/10^{30}\ \rm{erg\ s^{-1}})=a+b*\log(\dot{m}_{\rm jet}/10^9\ 
\rm{g\ s^{-1}})$.
\end{minipage}
\end{table*}

\clearpage

\begin{figure}
\plotone{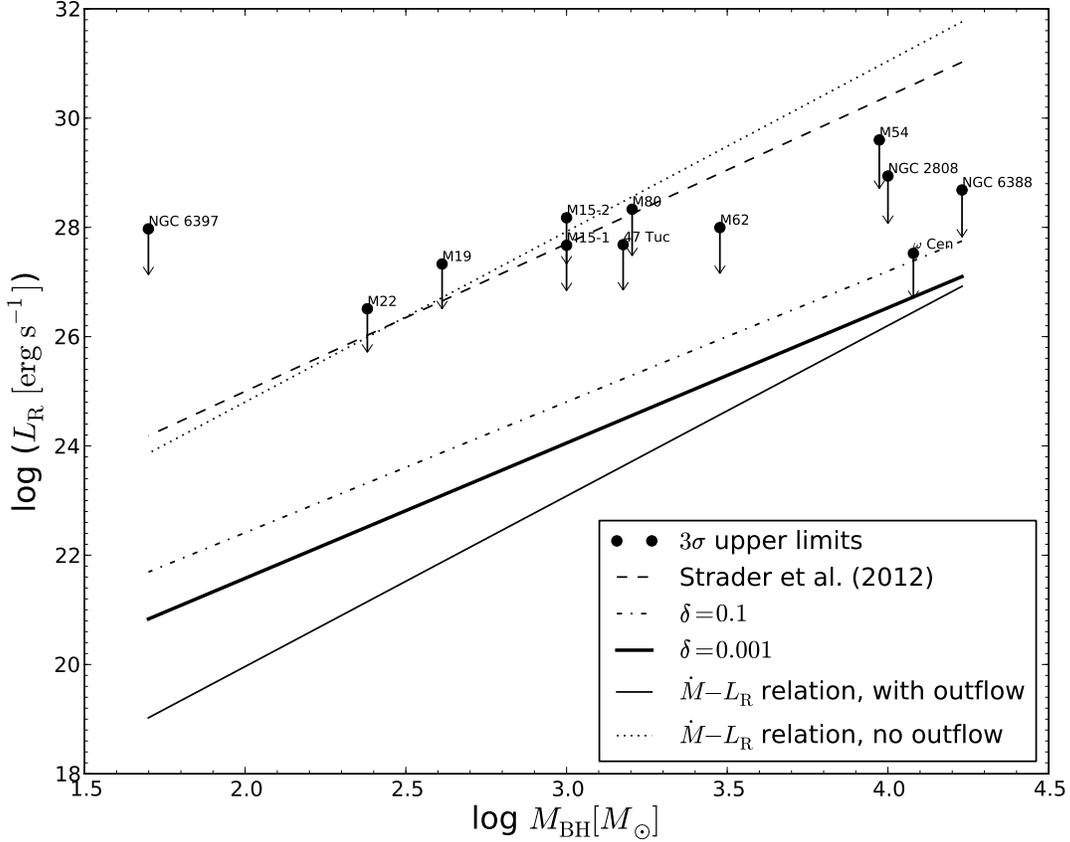}
\caption{
Radio luminosity $L_{\rm R}$ as a function of black hole mass $M_{\rm BH}$.
The filled circles represent the $3\sigma$ upper limit of $L_{\rm R}$
for each source, where the black hole masses are constrained via dynamical
modelings. The dashed line represents the predicted $L_{\rm R}$ as a
function of $M_{\rm BH}$ according to \cite{strader12}. The dot-dashed line and 
the thick solid line correspond to our new results with $\delta = 0.1$ and 
$0.001$, respectively. The thin solid line and the dotted line represent $L_{\rm R}$ 
estimated by the $\dot{M}-L_{\rm R}$ relation instead of the Fundamental Plane relation
with and without outflows, respectively.
}
\label{f1}
\end{figure}

\clearpage

\begin{figure}
\plotone{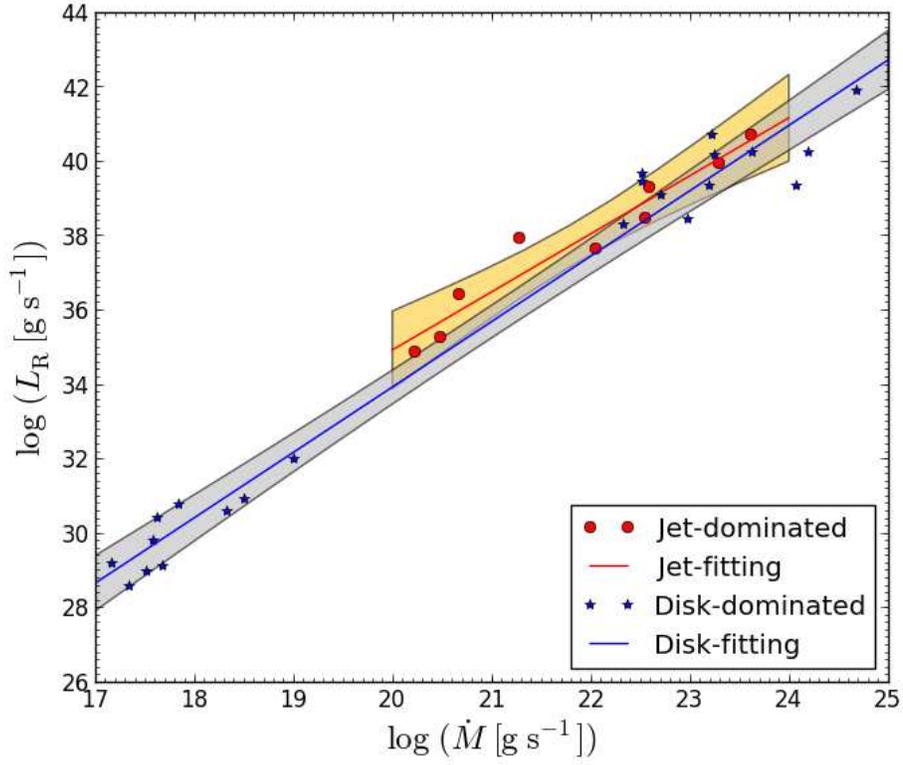}
\caption{
Relationship between accretion rate $\dot M$ and radio luminosity $L_{\rm R}$. 
The red circles (data) and the red line (fitting results) are for the sources
whose X-ray emission is dominated by the ADAF. The blue stars (data) and the blue 
line (fitting results) correspond to the sources whose X-ray emission mainly 
comes from the jet. The filled regions are for the $2\sigma$ confidence bands
of the fitting results.
}
\label{f2}
\end{figure}

\end{document}